\documentclass[a4paper,USenglish]{lipics-v2019}

\usepackage{amsmath,amssymb,amsthm}
\usepackage[ruled]{algorithm}
\usepackage{graphicx}
\usepackage{tikz}
\usetikzlibrary{plotmarks}
\usetikzlibrary{arrows.meta}
\usetikzlibrary{arrows.new}
\usepackage{pgfplots}
\usepackage{soul}
\usepackage{clrscode3e}
\usepackage{xparse}
\usepackage{booktabs}

\let\geq\geqslant
\let\setminus\smallsetminus

\newcommand{\brac}[1]{{\left(#1\right)}}
\newcommand{\set}[1]{\left\{#1\right\}}

\newcommand{\floor}[1]{{\left\lfloor #1 \right\rfloor}}
\newcommand{\ceil}[1]{{\left\lceil #1 \right\rceil}}

\newcommand{\Oh}[1]{O\brac{#1}}

\floatname{algorithm}{Listing}
\NewDocumentEnvironment{listing}{ o m }{
    \IfNoValueTF{#1}{
        \begin{algorithm}
            \caption{#2}
        \vspace{-0.7\baselineskip}
        \begin{codebox}
    }{
        \begin{algorithm}[#1]
            \caption{#2}
        \vspace{-0.7\baselineskip}
        \begin{codebox}
    }
}{
    \end{codebox}
    \vspace{-\baselineskip}
    \end{algorithm}
}

\renewcommand{\ElseIf}{\kill\addtocounter{indent}{-1}\liprint\textbf{else if} }
\newcommand{\Continue}{\textbf{continue}}
\newcommand{\Store}{\textbf{store }}
\newcommand{\Retrieve}{\textbf{retrieve }}

\title{An in-place, subquadratic algorithm\protect\\for permutation inversion}
\titlerunning{An in-place, subquadratic algorithm for permutation inversion} 

\author{G. Guśpiel}{Theoretical Computer Science Department, Faculty of Mathematics and Computer Science, Jagiellonian University, Kraków, Poland}{guspiel@tcs.uj.edu.pl}{https://orcid.org/0000-0002-3303-8107}{}

\authorrunning{G. Guśpiel}

\Copyright{Grzegorz Guśpiel}



\category{}

\supplement{}

\funding{This work was partially supported by the Polish Ministry of Science and Higher Education grant DI2013 000443.}

\acknowledgements{I would like to thank Grzegorz Gutowski
for his enormous help with the preparation of this manuscript.
I also thank prof.~Paweł Idziak for many valuable comments,
Bartłomiej Bosek for introducing me to this problem and 
Vladan Majerech for pointing out a mistake in the previous version of this manuscript.
            }

\nolinenumbers 

\hideLIPIcs  

\EventEditors{John Q. Open and Joan R. Access}
\EventNoEds{2}
\EventLongTitle{42nd Conference on Very Important Topics (CVIT 2016)}
\EventShortTitle{CVIT 2016}
\EventAcronym{CVIT}
\EventYear{2016}
\EventDate{December 24--27, 2016}
\EventLocation{Little Whinging, United Kingdom}
\EventLogo{}
\SeriesVolume{42}
\ArticleNo{23}

\begin{document}

\maketitle

\begin{abstract}
    We assume the permutation $\pi$ is given by an $n$-element array in which the $i$-th element
    denotes the value $\pi(i)$.
    Constructing its inverse 
    in-place (i.e.~using $O(\log{n})$ bits of additional memory) 
    can be achieved in linear time with a simple algorithm. 
    Limiting the numbers that can be stored in our array to the range $[1...n]$ still allows a straightforward $O(n^2)$ time solution.
    The time complexity can be improved using randomization, but 
    this only improves the expected, not the pessimistic running time.
    We present a deterministic algorithm that runs 
    in $O(n^{3/2})$ time.
\end{abstract}

\section{Problem statement and previous work}

In the permutation inversion problem, the algorithm is given
a positive integer $n$ and
a permutation $\pi$ of the set $V = \set{1, ..., n}$
presented in an array $t$, where $t[i] = \pi(i)$ for all $i \in V$.
The goal is to 
perform a sequence of modifications of $t$,
so that eventually $t[i] = \pi^{-1}(i)$ for all $i \in V$.
In this paper 
we focus on algorithms that use
$\Oh{\log{n}}$ bits of additional memory.
Algorithms with this tiny bound on additional memory are called in-place algorithms.
The best known in-place deterministic algorithm for the permutation inversion problem runs in $\Oh{n^2}$.
            We reduce the running time to $\Oh{n^{3/2}}$.

\section{Previous work}

The in-place permutation inversion problem was considered in Knuth \cite{Knuth97},
where two solutions are described,
            by Huang and by Boothroyd.
These algorithms, however, 
are allowed to store any value from the range
$[-n, ..., n]$ in the array $t$.
This seems to bypass the heart of the problem.
In fact, the signs of the values in $t$
can be used as a vector of $n$~bits.
The problem is trivial when such a vector is allowed.
Algorithms described in~this paper
are allowed to store values only from the range $[1,...,n]$ in $t$
(which may lead to $t$ temporarily not representing a permutation).

First, we describe an $\Oh{n^2}$ time in-place algorithm. 
Observe that it is straightforward to reverse a single cycle of $\pi$:

\begin{listing}[H]{Pseudocode for reversing a cycle}
\Procname{$\proc{Reverse-Cycle}(\id{start})$}
\li     $\id{cur} \gets t[\id{start}]$
\li     $\id{prev} \gets \id{start}$
\li     \While $\id{cur} \neq \id{start}$
\li         \Do
                $\id{next} \gets t[\id{cur}]$
\li             $t[\id{cur}] \gets \id{prev}$
\li             $\id{prev} \gets \id{cur}$
\li             $\id{cur} \gets \id{next}$
            \End
\li     $t[\id{start}] \gets \id{prev}$
\end{listing}

For a cycle in the permutation,
let its \emph{leader} be the smallest element in this cycle.
The following code finds in $\Oh{n}$ time the leader
of the cycle (containing a given element  \id{start}):

\begin{listing}[H]{Pseudocode for finding the leader of a cycle}
\Procname{$\proc{Cycle-Leader}(\id{start})$}
\li     $\id{cur} \gets t[\id{start}]$
\li     $\id{smallest} \gets \id{start}$
\li     \While $\id{cur} \neq \id{start}$
\li         \Do
                $\id{smallest} \gets \min(\id{smallest}, \id{cur})$
\li             $\id{cur} \gets t[\id{cur}]$
            \End
\li     \Return \id{smallest}
\end{listing}

To obtain the inverse of $\pi$, it suffices to reverse each of the cycles exactly once:

\begin{listing}[H]{A quadratic in-place algorithm for permutation inversion}\label{lst:quadalgo}
\li     \For $i \gets 1$ \To $n$
\li         \Do
                \If $\proc{Cycle-Leader}(i) \isequal i$  \label{li:is-leader}
\li                 \Then $\proc{Reverse-Cycle}(i)$
                \End
            \End
\end{listing}

In 1995, Fich, Munro and Poblete~\cite{Fich95} published a paper on a similar topic:
the permutation $\pi$ is
given by means of an oracle and the goal is to permute the contents of an array according to $\pi$.
They provided an algorithm with running time $\Oh{n \log^2{n}}$ that uses
$\Oh{\log^2{n}}$ bits of additional memory.
The concept of a cycle leader comes from this paper.

In 2015, the methods of \cite{Fich95} were extended to the permutation inversion problem by El-Zein, Munro and Robertson \cite{ElzeinMR2016,Robertson15},
who gave an algorithm with running time $\Oh{n \log{n}}$ 
that uses $\Oh{\log^2{n}}$ bits of additional memory.

It is interesting to note that the quadratic algorithm
can be easily modified to achieve $\Oh{n \log{n}}$ expected running time.
In \cite{Fich95}, the authors point out, attributing the idea to
a personal communication with Impagliazzo,
that one can use a randomly chosen hash function $h$ and choose the cycle leader to be the element $i$
with the smallest value $h(i)$.
This idea can be applied to cycle inversion as well.
When visiting element $i$ in the main loop,
we start reversing its cycle and either complete this operation
or encounter an element $j$ such that $h(j)<h(i)$,
in which case we revert the operation.
This way, assuming no hash collisions, every cycle is reversed once and 
if $h$ is chosen randomly,
the average time spent 
in the main loop at any single element is $\Oh{\log{n}}$.
Thus, we obtain an in-place algorithm with expected running time $\Oh{n \log{n}}$.

To our knowledge, we present the first subquadratic deterministic in-place
algorithm.
We compare it with previously known algorithms in Table~\ref{tab:perm}.
The running time of this algorithm is $\Oh{n^{3/2}}$.
Such time reduction is possible due to an alternative
representation of permutation cycles.

\begin{table}[h]
    \raggedleft
    \footnotesize
    \def\arraystretch{1.5}
    \begin{tabular}{l l l}
        \toprule[1.5pt]
        time complexity & required bits of additional memory & source \\
        \midrule
        $\Oh{n}$ & $\Oh{\log{n}}$ with $t[i] \in [-n, ..., n]$ & Huang \cite{Knuth97} \\
        $\Oh{n}$ & $\Oh{\log{n}}$ with $t[i] \in [-n, ..., n]$ & Boothroyd \cite{Knuth97} \\
        $\Oh{n^2}$ & $\Oh{\log{n}}$& folklore (Listing~\ref{lst:quadalgo}) \\
        $\Oh{n \log{n}}$ & $\Oh{\log^2{n}}$ & El-Zein, Munro, Robertson \cite{ElzeinMR2016,Robertson15} \\
        $\Oh{n \log{n}}$ in expectation & $\Oh{\log{n}}$ & Impagliazzo \cite{Fich95} \\
        $\Oh{n^{3/2}}$ & $\Oh{\log{n}}$ & this paper (Listing~\ref{lst_invert_2}) \\
        \bottomrule[1.5pt]
    \end{tabular}
    \caption{A summary of permutation inversion algorithms.} \label{tab:perm}
\end{table}

\section{A summary of the $\Oh{n^{3/2}}$ algorithm}

Observe that 
the quadratic algorithm runs in time $\Oh{n^{3/2}}$
on any instance in which each cycle is of size at most $\Oh{\sqrt{n}}$.
The complexity is worse when a large proportion of the elements
belong to cycles of greater sizes.
We modify the quadratic algorithm
to handle large cycles differently,
while keeping its behavior for small cycles.

During the course of the algorithm,
with the current state of array $t$ we associate a~directed graph $G_t$ on the vertex set $V$
and with edge $(i, t[i])$
for each $i \in V$.
The graph $G_t$ 
may contain loops and when we consider cycles,
loops are counted among them.
Initially, $G_t$ is a disjoint union of cycles.
When we modify $t$, we can get $G_t$ to be an~arbitrary graph
with outdegree of each vertex equal to~1.

We will introduce an alternative way of storing a long cycle in the array $t$.
This alternative representation of a cycle is achieved by redirecting
a small number of edges.
The result is a graph with 
multiple connected components;
each component
has $\Oh{\sqrt{n}}$ elements and is
a directed path leading to a cycle.
We will use
the lengths of these paths and cycles
to encode some information.
This information will allow us
to revert the edge redirections and obtain the original cycle.

When we first encounter a long cycle, we reverse it and 
convert the reversed cycle to the alternative representation.
Next, whenever in the main loop we visit another vertex of the cycle,
we traverse only the component of the vertex,
which takes time $\Oh{\sqrt{n}}$.
Previously this step could take $\Oh{n}$
and this is the improvement that allows us to achieve $\Oh{n^{3/2}}$ time complexity.
When the loop finishes, all short cycles are reversed
and all the long cycles are stored using the alternative representation
with minor modifications.
Next, we perform one pass over all vertices to remove the modifications
and another pass to
convert long cycles
back from the alternative representation.

\section{Segments and cycle detection}

We call each connected component of the alternative representation to be a~\emph{segment}.
A~segment is a disjoint union of two directed graphs: a cycle and a path,
together with an edge from the last vertex of the path to one of the vertices in the cycle.
The first vertex of the path is called the \emph{beginning} of the segment,
the number of vertices in the segment is called its \emph{size},
and the path is called its \emph{tail}.

We use the sizes of the segments, as well as the sizes of their cycles,
to store some information.
Thus, whenever in the main loop variable $i$ becomes a vertex of a segment,
we need to be able to 
compute in-place the size of the cycle and the distance from $i$ to the cycle.
The problem is called cycle detection and
at least two different algorithms are known to solve it
in-place and in time proportional to the size of the segment:
the Floyd's cycle-finding algorithm (also called the tortoise and hare algorithm)
mentioned in Knuth \cite[Section 3.1, Exercise 6]{Knuth97-2},
and the Brent's \cite{Brent80} algorithm.
In Listing~\ref{lst_floyds}, we shortly present the first one.
The idea is as follows.
First, we initialize two variables, the \emph{tortoise} and the \emph{hare},
to point at element $i$ (called \id{start} in the following pseudocode).
Next, we simultaneously progress both variables:
the tortoise moves one step at a time and the hare moves two steps at a time,
where by a step we mean setting $v \gets t[v]$.
We stop when both variables point at the same element (which must happen after a number of steps 
that is linear in the number of vertices reachable from $i$).
Next, we bring the hare back to element $i$
and start progressing the pointers again, this time both by one step at a time.
They first meet at the beginning of the cycle, i.e.~the only vertex with indegree 2
in the segment,
which gives us the distance from $\id{start}$ to the cycle.
Finally, we use the tortoise one last time to compute the size of the cycle.
\begin{listing}{The Floyd's cycle-finding algorithm}\label{lst_floyds}
\Procname{$\proc{Tortoise-and-Hare}(\id{start})$}
\li     $\id{tortoise} \gets \id{hare} \gets \id{start}$
\li     $\id{cycle\_length} \gets \id{dist\_to\_cycle} \gets 0$
\li     \Repeat                                     \label{li:th-loop-begin}
\li         $\id{tortoise} \gets t[\id{tortoise}]$
\li         $\id{hare} \gets t[t[\id{hare}]]$
\li     \Until $\id{tortoise} \isequal \id{hare}$          \label{li:th-loop-end}
\li     $\id{hare} \gets \id{start}$
\li     \Repeat
\li         $\id{tortoise} \gets t[\id{tortoise}]$
\li         $\id{hare} \gets t[\id{hare}]$
\li         $\id{dist\_to\_cycle} \gets \id{dist\_to\_cycle} + 1$
\li     \Until $\id{tortoise} \isequal \id{hare}$
\li     \Repeat
\li         $\id{tortoise} \gets t[\id{tortoise}]$
\li         $\id{cycle\_length} \gets \id{cycle\_length} + 1$
\li     \Until $\id{tortoise} \isequal \id{hare}$
\li     \Return $(\id{cycle\_length}, \id{dist\_to\_cycle})$
\end{listing}

In our setting, it will be important to search for cycles
that are in a bounded distance from \id{start}
and have a bounded size.
In such a case, we would like $\proc{Tortoise-and-Hare}$ to perform a number of steps that is linear
in the sum of these two bounds.

Observe that if there is a cycle with distance at most $d$ from \id{start}
and size at most~$s$,
the loop in Listing~\ref{lst_floyds} in lines \ref{li:th-loop-begin} -- \ref{li:th-loop-end}
finishes after at most $d + s$ steps.
Therefore, if we modify \proc{Tortoise-and-Hare} to 
take a bound \id{max} and
return $(\const{nil},\const{nil})$ 
if this loop performs more than $\id{max}$ steps,
we obtain an algorithm that correctly finds the distance from \id{start}
to the cycle and the size of the cycle if their sum is at most \id{max},
and otherwise either successfully finds these values or returns $(\const{nil}, \const{nil})$.
This algorithm has the important property of running in $\Oh{\id{max}}$.
We present it in Listing~\ref{lst_floyds_limited}.
\begin{listing}{The Floyd's cycle-finding algorithm performing a bounded number of steps}\label{lst_floyds_limited}
\Procname{$\proc{Limited-Tortoise-and-Hare}(\id{start},\id{max})$}
\li     $\id{tortoise} \gets \id{hare} \gets \id{start}$
\li     $\id{cycle\_length} \gets \id{dist\_to\_cycle} \gets 0$
\li     \Repeat                                     
\li         \If $\id{max} \isequal 0$ 
\li             \Then
                    \Return $(\const{nil},\const{nil})$
            \End
\li         $\id{tortoise} \gets t[\id{tortoise}]$
\li         $\id{hare} \gets t[t[\id{hare}]]$
\li         $\id{max} \gets \id{max} - 1$
\li     \Until $\id{tortoise} \isequal \id{hare}$          
\li     $\id{hare} \gets \id{start}$
\li     \Repeat
\li         $\id{tortoise} \gets t[\id{tortoise}]$
\li         $\id{hare} \gets t[\id{hare}]$
\li         $\id{dist\_to\_cycle} \gets \id{dist\_to\_cycle} + 1$
\li     \Until $\id{tortoise} \isequal \id{hare}$
\li     \Repeat
\li         $\id{tortoise} \gets t[\id{tortoise}]$
\li         $\id{cycle\_length} \gets \id{cycle\_length} + 1$
\li     \Until $\id{tortoise} \isequal \id{hare}$
\li     \Return $(\id{cycle\_length}, \id{dist\_to\_cycle})$
\end{listing}

\section{The alternative representation of long cycles}

In this section, we define the alternative representation
of a long cycle
and show how to compute it.
Let $k = \ceil{\sqrt{n}}$.
Up to isomorphism, there are $k^2 \geq n$ different segments
 of size between $k + 1$ and $2k$
and of cycle length between 1 and $k$.
We choose different segments of the above form
to represent different elements of $V$.
This way, a segment can `store' a pointer to a vertex
-- we say that such a segment, or a pair of a segment size and a cycle length
of the above form, \emph{encodes} a vertex.
The encoding is the following bijection from $V$ to a subset of $\set{k+1,...,2k} \times \set{1,...,k}$: 
$\proc{Encode}(v) = (\floor{(v - 1) / k} + (k + 1), ((v - 1) \bmod k) + 1)$.
Its inverse is the function
$\proc{Decode}(s,c) = (s - (k + 1))k + c + 1$.

\begin{figure}[h]
    \center
    \input{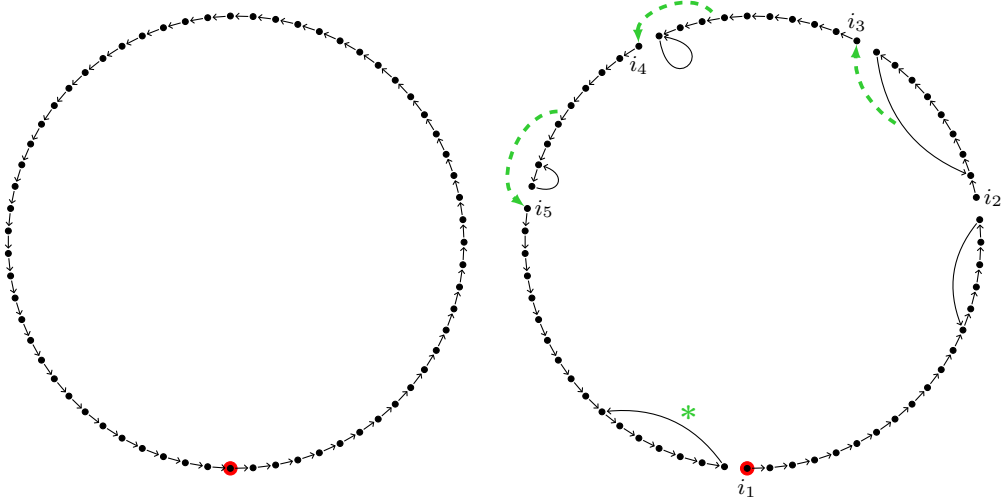}
    \caption{
    \label{fig_seg_repr}
        A long cycle $C$ and a possible segmentation of $C$, where $k = 8$ and $q=5$. The red color marks the cycle leader,
    a green line from a segment $X$ to a vertex $v$ indicates that $X$ encodes $v$,
    a green asterisk indicates the segment of size $2k+1$.}
\end{figure}

An intuitive understanding of the alternative representation is that the cycle is split
into segments such that all except $\Oh{\sqrt{n}}$ edges are preserved
(condition (C1)),
the first segment begins at the leader, and
every segment encodes the beginning of the next segment,
except the first and the last one, which we need to handle differently.
This is illustrated in Figure \ref{fig_seg_repr}.

We set a threshold for a cycle to be handled using the alternative representation as follows:
a cycle is called \emph{long},
if it has at least $4k+3$ vertices,
otherwise it is called \emph{short}.

For a directed graph $G$ and a subset $X$ of its vertex set,
let $G[X]$ denote the subgraph of $G$ induced by $X$.
For $X=\set{x_1,...,x_q}$, we simply write $G[x_1,...,x_q]$.
Let $c_1$ be the leader of a long cycle $C = (c_1, ..., c_p)$
on vertex set $V' \subseteq V$,
let $R$ be a directed graph on $V'$
with the outdegree of each vertex equal to 1
and let $S$ be an integer from $[k+1, 2k]$.
We say that a pair $(R, S)$
is a \emph{segment representation} of $C$
if there exist $q \geq 2$ integers $i_1, ..., i_q$
such that $1 = i_1 < ... < i_q < p$ and:
\begin{enumerate}
    \item[(C1)] for every $i \in \set{1, ..., p - 1} \setminus \set{i_2 - 1, ..., i_q - 1}$,
        the graph $R$ contains the edge $(c_i, c_{i+1})$;
    \item[(C2)] the graph $R[c_1, ..., c_{i_2 - 1}]$
        is a segment with beginning $c_1$,
            size between $2k+2$ and $4k+1$
            and cycle length $y$ such that
                $(S, y) = \proc{Encode}(c_{i_2})$;
            \item[(C3)] for every $j = 2, ..., q - 1$, the graph
        $R[c_{i_j}, c_{i_j + 1}, ..., c_{i_{j+1} - 1}]$
        is a segment with
            beginning $c_{i_j}$,
            size $x$
            and cycle length $y$ 
        such that $(x,y) = \proc{Encode}(c_{i_{j+1}})$;
    \item[(C4)] the graph $R[c_{i_q}, ..., c_p]$ is 
        a segment with
            beginning $c_{i_q}$,
            size $2k+1$
            and cycle length at most $k$.
\end{enumerate}
The graph $R$ is called a \emph{segmentation} of $C$.

The intuition behind this definition is as follows.
We want the segment representation of $C$ to store enough information
to be able to restore $C$.
Condition (C3) guarantees that
every segment except the first and the last one
encodes the beginning of the next one.
For the first segment, the beginning of its successor
can be decoded from 
the number $S$
and the length of the cycle in the segment
-- this is condition (C2).
The last segment 
has size $2k+1$,
while the length of the cycle can be any integer from $[1, ..., k]$
(condition (C4)).
The size $2k+1$ is special,
as no other segment can have this size,
and indicates that the segment is the last one.
The fact that we are free to choose the length of the cycle
in the last segment
is important and will be used later in the paper.

Recall that for a long cycle, we want to obtain a segment representation of its inverse.
In Listing~\ref{lst_make_seg}, we present a procedure that achieves this goal in-place and in time
linear in the size of the cycle.
We require that when $\proc{Make-Segments}$ is called,
vertex \id{leader} is the leader of a long cycle $C$ that is a subgraph of $G_t$.
We claim that when the procedure ends,
the subgraph of $G_t$ induced by the vertices of $C$
together with the number $S$ returned by $\proc{Make-Segments}$
is a segment representation of the inverse of $C$.

Before describing the pseudocode, we make several remarks.
First, for readability, the code in~Listing~\ref{lst_make_seg}, and others that follow,
do not necessarily
meet the in-place memory requirements if understood literally,
but it is straightforward to rewrite them in such a way that they do.
Second, 
we use the notation $t^i[v]$ defined as follows:
$t^0[v] = v$ and $t^i[v] = t[t^{i-1}[v]]$ for $i = 1, 2, \ldots $.
Finally, we note that except creating appropriate segments,
the procedure returns the value $\id{bg\_of\_sg\_created\_first}$
that is to be used later.
In fact, this variable keeps the beginning of the first segment
created by this procedure.

\begin{listing}{Producing a segment representation of the inverse of a cycle}\label{lst_make_seg}
\Procname{$\proc{Make-Segments}(\id{leader})$}
\li     $\id{to\_encode} \gets \id{leader}$
\li     $v_1 \gets t[\id{leader}]$
\li     \While \id{leader} is not among $v_1, t[v_1], t^2[v_1], ..., t^{2k}[v_1]$
\li         \Do
                \If $\id{to\_encode} \isequal \id{leader}$
\li                 \Then 
                        $s \gets 2k+1$
\li                     $\id{cycle\_length} \gets 1$
                        \>\>\>\> \Comment we can choose any length between 1 and $k$
\li                     $\id{bg\_of\_sg\_created\_first} \gets t^{2k}[v_1]$
\li                 \Else 
                        $(s, \id{cycle\_length}) \gets \proc{Encode}(\id{to\_encode})$
                \End
\li             let $v_i = t^{i-1}[v_1]$ for $i = 2, ..., s$
\li             $\id{next\_{v_1}} \gets t[v_s]$
\li             set $t[v_i] \gets v_{i-1}$ for $i = 2, ..., s$ \label{li:reverse-edges} 
\li             $t[v_1] \gets v_{\id{cycle\_length}}$ \label{li:set-cycle-length}
\li             $\id{to\_encode} \gets v_s$
\li             $v_1 \gets \id{next\_{v_1}}$
            \End
\li     $S \gets s$
\li     let $p$ be the smallest $i$ such that $t^{i-1}[v_1] = \id{leader}$
\li     let $v_i = t^{i-1}[v_1]$ for $i = 2, ..., p$
\li     $t[v_1] \gets \id{to\_encode}$ \label{li:attach}
\li     set $t[v_i] \gets v_{i-1}$ for $i = 2, ..., p$ \label{li:reverse-last-tail}
\li     \Return $(\id{bg\_of\_sg\_created\_first}, S)$
\end{listing}

In Listing~\ref{lst_make_seg}, we simultaneously reverse the cycle and form new segments.
We maintain  positions $v_1$ and \id{to\_encode} in this cycle
such that $\id{to\_encode}$ is the predecessor of $v_1$.
Initially, $v_1$ is set to $t[\id{leader}]$ and $\id{to\_encode}$ to \id{leader}.
In each iteration of the loop, we form a new segment with beginning $t^{s-1}[v_1]$,
size $s$ and
vertices $v_1, t[v_1], ..., t^{s-1}[v_1]$,
and set $\id{to\_encode}$ to the beginning of this segment
and $v_1$ to next vertex of the cycle.
The first segment created in the loop is made to satisfy condition (C4),
i.e.~just to have the special size $2k+1$.
Every next segment formed in this loop is created with size and cycle length
to encode the vertex \id{to\_encode}, as required by (C3).
Line \ref{li:reverse-edges} is responsible for reversing the edges and line \ref{li:set-cycle-length} for setting appropriate cycle length.
The loop condition ensures that even if $s$ is set to the maximum possible value,
i.e.~$2k+1$, the newly formed segment will not contain \id{leader}.
After the loop finishes, the negation of the loop condition
guarantees that the path $(v_1, t[v_1], t^2[v_1], ..., \id{leader})$
has at most $2k+1$ vertices.
In lines \ref{li:attach}--\ref{li:reverse-last-tail}, this path is reversed and attached to the last created segment.
We note that because long cycles are defined to have at least $4k+3$ vertices,
at least two segments are created, as required in the definition of the segment representation.
Thus, the last created segment has size at most $2k$ before
attaching the path, and at most $4k+1$ after attaching it.
Next, as the loop condition is true before the last iteration,
the last created segment after the attachment of the path has at least $2k+2$ vertices.
Together with the fact that its cycle length and the value $S$ are
appropriately set, we get that the condition (C2) is satisfied.

Now consider the problem of restoring the original cycle from its segment representation.
We start at the leader, 
decode the beginning of the next segment using the value $S$,
redirect a single edge and proceed to the next segment.
This time, we use the size of the segment and the size of its cycle
to decode the beginning of the next segment.
We repeat this step several times
until we encounter the segment of size $2k+1$.
Then we redirect the last edge to the cycle leader (stored in a separate variable) and stop.
According to condition (C4),
the cycle in this last encountered segment may have any length
between 1 and $k$.
The code actually returns this length,
as it is used by our inversion algorithm to store additional information.
The full procedure to restore a long cycle is presented in Listing~\ref{lst_restore}.

\begin{listing}{Restoring a long cycle from its segment representation}\label{lst_restore}
\Procname{$\proc{Restore-Long-Cycle}(\id{leader}, S)$}
\li     $(\id{cycle\_length}, \id{dist\_to\_cycle}) \gets \proc{Tortoise-and-Hare}(\id{leader})$
\li     $\id{segment\_size} \gets \id{dist\_to\_cycle} + \id{cycle\_length}$
\li     $\id{last} \gets t^{\id{segment\_size}-1}[\id{leader}]$
\li     $\id{bg} \gets t[\id{last}] \gets \proc{Decode}(S, \id{cycle\_length})$
\li     \While \const{true}
\li         \Do
                $(\id{cycle\_length}, \id{dist\_to\_cycle}) 
                    \gets \proc{Tortoise-and-Hare}(\id{bg})$
\li             $\id{segment\_size} \gets \id{dist\_to\_cycle} + \id{cycle\_length}$
\li             $\id{last} \gets t^{\id{segment\_size}-1}[\id{bg}]$
\li             \If $\id{segment\_size} \isequal 2k+1$
\li                 \Then
                        $t[\id{last}] \gets \id{leader}$
\li                     \Return \id{cycle\_length}
\li                 \Else
                        $\id{bg} \gets t[\id{last}] \gets \proc{Decode}(\id{segment\_size}, \id{cycle\_length})$
                    \End
            \End
\end{listing}

\section{The $\Oh{n^{3/2}}$ algorithm}

Let us summarize how we change the $\Oh{n^2}$ algorithm to obtain
$\Oh{n^{3/2}}$ running time.
First, when a long cycle is first visited, it is reversed and converted into its segment representation.
The value $S$ is stored in another part of the graph $G_t$ -- this step is explained in Section \ref{sec:the_fix}, for now assume that this is somehow implemented, in appropriate time and memory complexity.
Short cycles are treated as before
and we add a third case to the main loop: if $i$ belongs to a tail of a segment,
we do nothing.
As a~result, after the main loop is complete,
all short cycles are reversed
and for every long cycle there is a segmentation of its inverse
with the cycles in all segments reversed
(notice that for every segment, its cycle $\sigma$ is reversed exactly once --
when $i$ becomes the leader of $\sigma$).

It remains to reverse the cycles in all segments again and then
restore the original cycle for every segment representation of a long cycle.
This is achieved with two more passes over all vertices
-- first we reverse the cycles in segments
and then we restore the long cycles.
The whole algorithm is shown in~Listing~\ref{lst_invert}.
We claim that it
inverts the permutation in-place in $\Oh{n^{3/2}}$ time.

\begin{listing}{The in-place $\Oh{n^{3/2}}$ time algorithm to invert a permutation}\label{lst_invert}
\Procname{$\proc{Invert-Permutation}()$}
\li     \For $i \gets 1$ \To $n$                                \label{li:first-loop-begin} 
\li         \Do
                $(\id{cycle\_length}, \id{dist\_to\_cycle}) 
                    \gets \proc{Limited-Tortoise-and-Hare}(i, 4k + 2)$ 
                    \label{li:first-call-to-limited-th}
\li             \If $(\id{cycle\_length},\id{dist\_to\_cycle}) \isequal (\const{nil},\const{nil})$
\li                 \Then
                        \Comment $i$ belongs to a long cycle \label{li:comment-long-cycle}
\li                     $(\id{bg\_of\_sg\_created\_first}, S) \gets \proc{Make-Segments}(i)$
                            \label{li:call-to-make-segments}
\li                     \Store $S$
\li             \ElseIf $\id{dist\_to\_cycle} \geq 1$
\li                 \Then 
                        \Comment $i$ belongs to a tail of a segment \label{li:comment-tail}
\li                     \Continue
\li             \ElseNoIf \Comment $i$ belongs to a short cycle \label{li:comment-short-cycle}
\li                 \If $\proc{Cycle-Leader}(i) \isequal i$ \label{li:run-cycle-leader}
\li                     \Then $\proc{Reverse-Cycle}(i)$ \label{li:run-reverse-cycle}
                    \End
                \End
            \End \label{li:first-loop-end}

\li     \For $i \gets 1$ \To $n$                                \label{li:second-loop-begin}
\li         \Do
                $(\id{cycle\_length}, \id{dist\_to\_cycle}) 
                    \gets \proc{Limited-Tortoise-and-Hare}(i,4k+2)$
                    \label{li:second-call-to-limited-th}
\li             \If $\id{dist\_to\_cycle} \isequal 1$       
\li                 \Then $\proc{Reverse-Cycle}(t[i])$
                \End
            \End                                                \label{li:second-loop-end}

\li     \For $i \gets 1$ \To $n$                                \label{li:third-loop-begin}
\li         \Do
                $(\id{cycle\_length}, \id{dist\_to\_cycle}) 
                    \gets \proc{Limited-Tortoise-and-Hare}(i, 4k+1)$
                    \label{li:third-call-to-limited-th}
\li             \If $\id{dist\_to\_cycle} \neq \const{nil}$ and $\id{dist\_to\_cycle} \geq 1$ 
                    \label{li:belongs-2}
                    \Then 
\li                     \Comment $i$ belongs to a tail of a segment
                                    and is the leader of a long cycle in $\pi$
\li                     \Retrieve $S$
\li                     $\proc{Restore-Long-Cycle}(\id{i},S)$
                            \label{li:restore}
                \End
            \End                                                \label{li:third-loop-end}
\end{listing}

Let us prove the correctness of the algorithm.
First, note that the call to $\proc{Limited-Tortoise-and-Hare}$ in line 
\ref{li:first-call-to-limited-th} returns $(\const{nil},\const{nil})$
if and only if
the vertex $i$ belongs to a~long cycle.
This is because we only create segments of size at most $4k+1$ and long cycles are defined to have at least $4k+3$ vertices.
Thanks to this observation, we know that our algorithm correctly determines
whether $i$ is a part of a tail, a short, or a long cycle,
as asserted in lines \ref{li:comment-long-cycle}, \ref{li:comment-tail}, \ref{li:comment-short-cycle}.
Second, note that the call to $\proc{Limited-Tortoise-and-Hare}$ in line
\ref{li:second-call-to-limited-th} never returns $(\const{nil},\const{nil})$,
    because at this point every component of $G_t$
is either a segment of size at most $4k+1$ or a short cycle.
Third, note that the call to $\proc{Limited-Tortoise-and-Hare}$ in line 
\ref{li:third-call-to-limited-th} does not return $(\const{nil},\const{nil})$
if $i$ belongs to a~tail of a segment, because every segment ever created has size at most $4k+1$.
Therefore, no tail is omitted by the loop in lines \ref{li:third-loop-begin} -- \ref{li:third-loop-end}.

Now, let $C$ be a cycle in $\pi$ with vertex set $V'$. 
If $C$ is short, our algorithm reverses it only once
-- when $i$ is the leader of $C$.
If $C$ is long, consider all the moments when our algorithm modifies $t[v]$ for $v \in V'$.
The first such moment is when during the loop in~lines~\ref{li:first-loop-begin}~--~\ref{li:first-loop-end}
vertex $i$ becomes the leader of $C$, and then the cycle is
reversed and split into segments.
Next, during the same loop,
all that happens is that every cycle in every segment is reversed exactly once
-- when $i$ becomes the leader of the cycle.
During the loop in lines \ref{li:second-loop-begin} -- \ref{li:second-loop-end},
every cycle in every segment is reversed again, also exactly once
(because for every such cycle $\sigma$ there is exactly one vertex
with distance to $\sigma$ equal to~1).
Thus, after this loop, the graph $G_t[V']$ is again a segmentation of the inverse of $C$.
Later, the first change to $G[V']$ that occurs in the loop in lines \ref{li:third-loop-begin} -- \ref{li:third-loop-end}
happens when $i$~becomes the leader of $C$.
Then, 
the line \ref{li:restore}
sets $G_t[V']$ to be the inverse of $C$.
From this moment on, the code does not modify $G_t[V']$.
This completes the correctness proof.

We now prove that the running time of \proc{Invert-Permutation} is $\Oh{n^{3/2}}$.
To see that the loop in lines \ref{li:first-loop-begin} -- \ref{li:first-loop-end}
runs in $\Oh{n^{3/2}}$,
note that:
\begin{itemize}
    \item the call to \proc{Limited-Tortoise-and-Hare} in line \ref{li:first-call-to-limited-th}
        takes time $\Oh{4k + 2} = \Oh{\sqrt{n}}$,
    \item the call to \proc{Make-Segments} in line \ref{li:call-to-make-segments}
        takes time proportional to the length of the cycle to which $i$ belongs,
        potentially $\Oh{n}$,
        but occurs just once for every long cycle,
    \item lines \ref{li:run-cycle-leader} -- \ref{li:run-reverse-cycle} take $\Oh{\sqrt{n}}$ time
        (at each iteration of the loop),
        as the cycle the vertex~$i$ belongs to is short.
\end{itemize}
As for the loop in lines \ref{li:second-loop-begin} -- \ref{li:second-loop-end},
  the instructions $\proc{Limited-Tortoise-and-Hare}(i, 4k+2)$ 
 and $\proc{Reverse-Cycle}(t[i])$ take time $\Oh{\sqrt{n}}$:
 the former because $\Oh{4k+2} = \Oh{\sqrt{n}}$, and the latter because the reversed cycle is a part
 of a segment of size $\Oh{\sqrt{n}}$.
Finally, in the loop in lines \ref{li:third-loop-begin} -- \ref{li:third-loop-end},
the call to~$\proc{Limited-Tortoise-and-Hare}$ takes time $\Oh{4k+1} $ $ = \Oh{\sqrt{n}}$,
and the call to $\proc{Restore-Long-Cycle}$ is $\Oh{n}$ but occurs exactly once for each long cycle in $\pi$.

As a final remark, we note that
the code can easily be rewritten to work in-place,
and thus our analysis of Listing~\ref{lst_invert} is complete.

\section{Storing the value $S$}\label{sec:the_fix}

Recall that the length of the cycle in the segment of size $2k+1$ of a segmentation
can be any integer between 1 and $k$.
We call this cycle the \emph{free cycle} of that segmentation
and we use it to store information.
We now make use of the return value of the procedure $\proc{Make-Segments}$.
The procedure returns a pair of integers:
the beginning of the segment
with the free cycle 
and the value $S$.

In our algorithm, there are two passes over all vertices.
In both passes, long cycles are visited in the same order,
say $C_1, ..., C_r$
(here each $C_i$ denotes a cycle before its reversal).
Let $C_1', ..., C_r'$ be the respective inverses of $C_1,...,C_r$.
After the first pass, in place of every $C_i$
there is a segment representation $(R_i,S_i)$ of $C_i'$.
The idea is to store $S_i$
as the length of the free cycle in $R_{i-1}$.
This way, when restoring $C_{i-1}'$, we can retrieve
the value $S_i$ and use it to restore $C_i'$.
This is done for $i \geq 2$,
while the value $S_1$ 
 is stored in
 a separate variable, named \id{first\_S}.
In~Listing~\ref{lst_invert_2}, we present the updated pseudocode of the $\Oh{n^{3/2}}$ time algorithm.
The only difference compared to the previous version is 
the implementation of the operations \Store and \textbf{retrieve}.

\begin{listing}{The in-place $\Oh{n^{3/2}}$ time algorithm to invert a permutation}\label{lst_invert_2}
\Procname{$\proc{Invert-Permutation}()$}
\li     $\id{storage}, \id{first\_S} \gets \const{nil}$
\li     \For $i \gets 1$ \To $n$                                
\li         \Do
                $(\id{cycle\_length}, \id{dist\_to\_cycle}) 
                    \gets \proc{Limited-Tortoise-and-Hare}(i, 4k + 2)$ 
\li             \If $(\id{cycle\_length},\id{dist\_to\_cycle}) 
                        \isequal (\const{nil},\const{nil})$ 
\li                 \Then
                        \Comment $i$ belongs to a long cycle 
\li                     $(\id{bg\_of\_sg\_created\_first}, S) \gets \proc{Make-Segments}(i)$
\li                     \If $\id{storage} \isequal \const{nil}$
\li                         \Then 
                                $\id{first\_S} \gets \id{S}$
\li                         \Else
                                set the length of the cycle in the segment beginning at $\id{storage}$ to \id{S}
                            \End
\li                     $\id{storage} \gets \id{bg\_of\_sg\_created\_first}$
\li             \ElseIf $\id{dist\_to\_cycle} \geq 1$
\li                 \Then 
                        \Comment $i$ belongs to a tail of a segment 
\li                     \Continue
\li             \ElseNoIf \Comment $i$ belongs to a short cycle 
\li                 \If $\proc{Cycle-Leader}(i) \isequal i$ 
\li                     \Then $\proc{Reverse-Cycle}(i)$ 
                    \End
                \End
            \End 

\li     \For $i \gets 1$ \To $n$                                
\li         \Do
                $(\id{cycle\_length}, \id{dist\_to\_cycle}) 
                    \gets \proc{Limited-Tortoise-and-Hare}(i,4k+2)$
\li             \If $\id{dist\_to\_cycle} \isequal 1$       
\li                 \Then $\proc{Reverse-Cycle}(t[i])$
                \End
            \End                                                

\li     $\id{S} \gets \id{first\_S}$
\li     \For $i \gets 1$ \To $n$                                
\li         \Do
                $(\id{cycle\_length}, \id{dist\_to\_cycle}) 
                    \gets \proc{Limited-Tortoise-and-Hare}(i, 4k+1)$
\li             \If $\id{dist\_to\_cycle} \neq \const{nil}$ and $\id{dist\_to\_cycle} \geq 1$ 
                    \Then 
\li                     \Comment $i$ belongs to a tail of a segment
                                    and is the leader of a long cycle in $\pi$
\li                     $S \gets \proc{Restore-Long-Cycle}(\id{i},S)$
                \End
            \End                                                
\end{listing}

We omit the correctness, memory consumption and time consumption analysis of~the~code,
as it is analogous to the analysis already done for Listing~\ref{lst_invert}.

\section{Suggestions for further research}

As we said earlier, the problem can be solved in $O(n\log{n})$ expected time
using a randomized algorithm.
Whether
there is a deterministic solution running in $O(n\log^c{n})$ for some constant $c$
seems to be an interesting question.

\bibliography{paper}{}

\end{document}